\documentclass[conference]{IEEEtran}
\IEEEoverridecommandlockouts

\usepackage{cite}
\usepackage{amsmath,amssymb,amsfonts}
\usepackage{algorithmic}
\usepackage{graphicx}
\usepackage{textcomp}
\usepackage{xcolor}
\usepackage{courier}
\usepackage{url}
\usepackage{colortbl}
\definecolor{lightgray}{gray}{0.95}
\usepackage{tabularx}
\usepackage{multirow} 
\usepackage{bm}
\usepackage{hyperref}
\usepackage{makecell}
\usepackage{booktabs} 
\usepackage{comment}
\usepackage{soul}
\usepackage{listings}

\def\BibTeX{{\rm B\kern-.05em{\sc i\kern-.025em b}\kern-.08em
    T\kern-.1667em\lower.7ex\hbox{E}\kern-.125emX}}
\begin{document}

\title{Toward an Intrusion Detection System for a Virtualization Framework in Edge Computing}

\author{\IEEEauthorblockN{Everton de Matos}
\IEEEauthorblockA{\textit{Secure Systems Research Center} \\
\textit{Technology Innovation Institute (TII)}\\
Abu Dhabi, Unied Arab Emirates \\
everton.dematos@tii.ae}
\and
\IEEEauthorblockN{Hazaa Alameri}
\IEEEauthorblockA{\textit{Secure Systems Research Center} \\
\textit{Technology Innovation Institute (TII)}\\
Abu Dhabi, Unied Arab Emirates \\
hazaa.alameri@tii.ae}
\and
\IEEEauthorblockN{Willian Tessaro Lunardi}
\IEEEauthorblockA{\textit{Secure Systems Research Center} \\
\textit{Technology Innovation Institute (TII)}\\
Abu Dhabi, Unied Arab Emirates \\
willian.lunardi@tii.ae}
\and
\IEEEauthorblockN{Martin Andreoni}
\IEEEauthorblockA{\textit{Secure Systems Research Center} \\
\textit{Technology Innovation Institute (TII)}\\
Abu Dhabi, Unied Arab Emirates \\
martin.andreoni@tii.ae}
\and
\IEEEauthorblockN{Eduardo Viegas}
\IEEEauthorblockA{\textit{Graduate Program in Computer Science} \\
\textit{Pontifical Catholic University of Paraná}\\
Curitiba, Brazil \\
eduardo.viegas@ppgia.pucpr.br}
}


\maketitle

\begin{abstract}
Edge computing pushes computation closer to data sources, but it also expands the attack surface on resource-constrained devices. This work explores the deployment of the Lightweight Deep Anomaly Detection for Network Traffic (LDPI) integrated as an isolated service within a virtualization framework that provides security by separation. LDPI, adopting a Deep Learning approach, achieved strong training performance, reaching AUC $\approx 0.999$ (5-fold mean) across the evaluated packet-window settings $(n,l)$, with high F1 at conservative operating points. We deploy LDPI on a laptop-class edge node and evaluate its overhead and performance in two scenarios: (i) comparing it with representative signature-based \mbox{IDSes} (Suricata and Snort) deployed on the same framework under identical workloads, and (ii) while detecting network flooding attacks.   
\end{abstract}

\begin{IEEEkeywords}
Edge Computing, Intrusion Detection System, Lightweight Deep Anomaly Detection, Network Security.
\end{IEEEkeywords}

\section{Introduction}
\label{sec:introduction}
The proliferation of the Internet of Things (IoT) devices and the increasing demand for real-time data processing have driven the adoption of Edge Computing \cite{SHI16}. Unlike traditional Cloud Computing, which relies on centralized data centers, Edge Computing performs data processing at or near the data source. This approach reduces latency, improves response times, and reduces the bandwidth required for data transmission to centralized servers \cite{MAO17}.

While Edge Computing offers numerous advantages, it also introduces new security challenges. Edge devices are often deployed in diverse and potentially insecure environments, making them vulnerable to various attacks \cite{RAN21}. The need for robust security measures is paramount to protect sensitive data and ensure the integrity and availability of Edge Computing systems \cite{ALI21}.

A significant challenge is the detection and mitigation of attacks. Traditional security measures, such as firewalls and Intrusion Detection Systems (IDS), must be adapted to edge environments' unique constraints and requirements. IDS specifically tailored for Edge Computing plays a crucial role in identifying and mitigating threats in real-time \cite{LIN18}. These systems monitor network traffic for suspicious activities, analyze data packets, and generate alerts or take preventive actions when potential threats are detected \cite{THA21}. Lightweight and efficient security solutions are essential to address these challenges without imposing significant overhead on edge devices \cite{TIB19}.

In this context, the Lightweight Deep Anomaly Detection for Network Traffic (LDPI) stands out by integrating Zero Trust Architecture principles to enhance the security of edge environments. LDPI was designed to detect and mitigate various security threats in real-time, providing a crucial layer of defense for Edge Computing systems. Using unsupervised and semi-supervised learning approaches, the LDPI monitors network interfaces to identify deviations from expected traffic patterns. It primarily relies on normal traffic data for training, though incorporating additional malicious samples can enhance the method's effectiveness. 

Virtualization platforms provide security by separation, isolating system components in minimally privileged virtual machines. It reduces the Trusted Computing Base (TCB), confines failures and compromises, and enables precise policy enforcement across trust domains. Building on this principle, we deployed LDPI as an isolated service within a virtualization framework, monitoring network flows while remaining decoupled from application workloads. To instantiate and evaluate this design, we tested LDPI on the Ghaf virtualization framework \cite{GHA24}. Ghaf is based on NixOS, which brings the benefits of reproducible builds, declarative configuration, and a strong emphasis on security \cite{DOL08}.

This paper evaluates the LDPI when deployed in a virtualization framework within a laptop platform, focusing on its overall performance in Edge Computing environments. The main contributions of this paper are:
\begin{itemize}
    \item Deployment of the LDPI module on a real-world edge environment, utilizing a virtualization framework as the supporting platform;
    \item Evaluation of LDPI’s performance in detecting and responding to security threats, focusing on metrics such as memory usage and CPU consumption;
    \item Comparative study against representative traditional \mbox{IDSes}, deployed on the same virtualization framework and exercised under identical workloads.
\end{itemize}

The remainder of this paper is organized as follows: Section \ref{sec:background} presents background information on concepts used by this work. Section \ref{sec:related} provides an overview of related work in the field of intrusion detection. Section \ref{sec:ldpi} details the LDPI, including its design and implementation within the Ghaf Framework. Section \ref{sec:eva} presents our experimental validation of LDPI. Finally, Section \ref{sec:conc} concludes the paper.

\section{Background}
\label{sec:background}
Virtualization is a pivotal technology in computing that allows multiple operating systems to run concurrently on a single physical hardware platform \cite{MORA15}. This is achieved through a hypervisor, which manages the distribution of hardware resources among different Virtual Machines (VMs). In embedded systems, virtualization is particularly important for ensuring efficient use of resources, enhancing system security, and providing robust isolation between different system functions.


Embedded systems often utilize virtualization, allowing multiple operating functions to be segregated and secured on a single physical machine \cite{TIB19} \cite{HEI08}. For instance, in automotive systems, virtualization can separate safety-critical functions from infotainment systems, ensuring that a failure in one does not affect the other. Similarly, in industrial automation, virtualization allows legacy systems to coexist with modern control software on the same hardware, facilitating upgrades and reducing costs.

The Ghaf Framework\footnote{\url{https://github.com/tiiuae/ghaf}} is designed to provide a modular and scalable edge device software architecture. It emphasizes virtualization and the integration of Zero Trust Architecture (ZTA) principles to enhance security and manageability. ZTA is a security model that eliminates implicit trust within a network, ensuring all access requests are thoroughly authenticated, authorized, and encrypted \cite{ZTA20}. 

Central to the Ghaf Framework is its modular approach, which uses virtualization technology to create multiple VMs that segregate system functions into isolated environments. This structure not only enhances security by reducing the TCB but also increases flexibility in system configuration, accommodating a wide range of applications from secure communications to advanced Edge Computing solutions, thereby leveraging the core principles of virtualization to optimize and secure edge device architecture \cite{TIB21}. Each VM is tailored to perform specific functions, contributing to a robust and secure architecture. These include an Admin VM for system management and policy enforcement, a Connection VM for handling networking tasks, a Display VM for graphical user interactions, and Storage VMs for data integrity and protection. Additionally, Application VMs run various user applications in isolated environments, reducing the risk of cross-contamination between processes and enhancing overall system security.

\section{Related Work}
\label{sec:related}

Suricata\footnote{\url{https://suricata.io/}} is an open-source IDS/IPS (Intrusion Prevention System) designed for deep packet inspection, network monitoring, and security event management. It operates as a rule-based IDS, utilizing predefined patterns and signatures to detect malicious activity within network traffic. Suricata is widely recognized for its ability to perform extensive protocol decoding and traffic inspection, providing detailed insights into network behavior and security events. Its architecture supports integration with broader security tools, making it a robust choice for traditional network security deployments.

Snort\footnote{\url{https://www.snort.org/}} is an open-source network IDS/IPS that uses a signature-based detection model. It performs deep packet inspection with protocol decoders and preprocessors to normalize traffic and uncover malicious patterns. Snort supports both passive IDS and inline IPS modes, offers a flexible rule language, and benefits from a broad ecosystem of community- and vendor-maintained signatures. It integrates readily with logging, SIEM, and packet-capture pipelines. Its strengths are precise detection of known attack patterns and policy enforcement, with effectiveness tied to the curation and freshness of its rule sets.

In contrast to Suricata and Snort, LDPI employs a deep learning-based anomaly detection approach, which enables it to identify previously unseen threats or anomalies that are not captured by signature-based systems like Suricata and Snort. Both systems share the ability to monitor real-time network traffic and integrate with logging and alerting frameworks, but their operational focus differs significantly. LDPI anonymizes sensitive traffic data, such as IP addresses, to prioritize privacy - a feature not inherently present in Suricata.

\section{Lightweight Deep Anomaly Detection for Network Traffic}
\label{sec:ldpi}
The Ghaf Framework, with its modular and scalable architecture, allows for the easy inclusion of specialized modules to boost its functionality. One such module, the Lightweight Deep Anomaly Detection for Network Traffic (LDPI)\footnote{\url{https://github.com/tiiuae/srta-ldpi}}, is integrated into the Connection VM of the Ghaf framework. Strategically placed, LDPI monitors and detects network traffic anomalies like DoS (Denial-of-Service), brute force, infiltration, and malware traffic. LDPI has a configurable option to analyze both incoming and outgoing traffic, providing adaptability according to network security needs.

\begin{figure}
\centering\includegraphics[width=.45\textwidth]{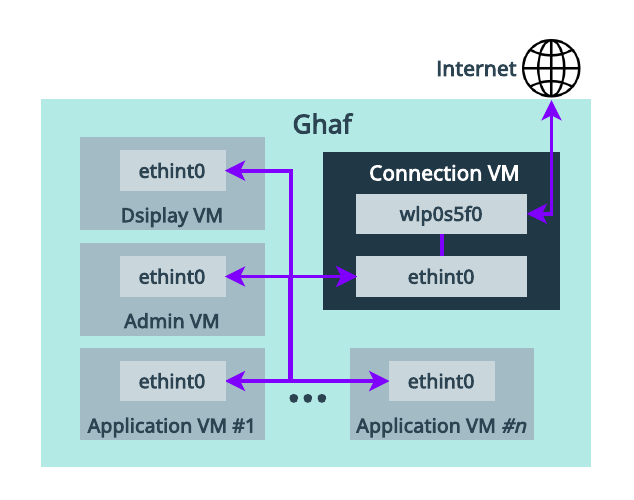}
\caption{\label{fig:net-vm-interfaces}Network connection architecture on Ghaf.}
\end{figure}

By design, Ghaf's Connection VM is equipped with two network interfaces: \textit{wlp0s5f0} and \textit{ethint0}, as illustrated in Figure \ref{fig:net-vm-interfaces}. The \textit{wlp0s5f0} interface is only present at the Connection VM and provides the primary link between the Ghaf framework and external networks. In contrast, each individual VM within Ghaf has a separate \textit{ethint0} network interface that enables internal communication and also makes possible for each VM to access the Internet through the Connection VM external interface. Given that \textit{wlp0s5f0} serves as the primary interface connecting Ghaf to the outside world, it is the logical and most effective location for deploying the LDPI module. This placement enables LDPI to monitor and analyze all inbound and outbound traffic flowing through Ghaf. We utilize \textit{iptables} to actively block IP addresses flagged as anomalous by LDPI. This approach automatically restricts suspicious network traffic, thus preventing potential threats from interacting with the system.

\begin{figure}
\centering\includegraphics[width=0.51\textwidth]{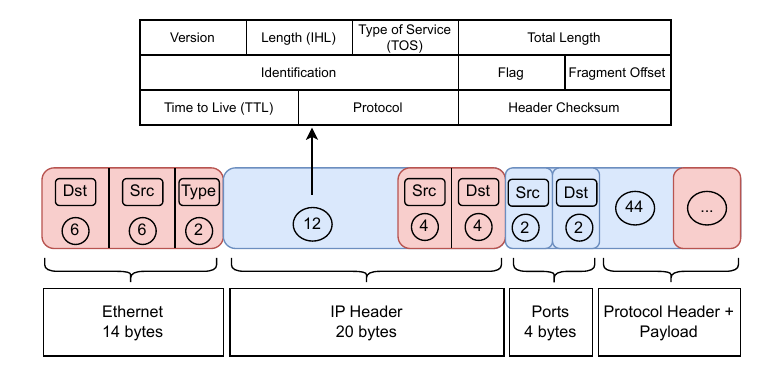}
\caption{\label{fig:ldpi_flow}Illustration of an IPv4 network flow analyzed by LDPI, highlighting ignored bytes (red) and analyzed bytes (blue).}
\end{figure}

LDPI, building on our previous work \cite{LUN23}, adopts a Deep Learning approach to detect anomalies in network traffic. It analyzes raw bytes from initial packets in new network flows. Figure \ref{fig:ldpi_flow} illustrates an example of an IPv4 network flow analyzed by LDPI. Bytes ignored during anomaly detection are highlighted in red, while bytes used for analysis are shown in blue. The first 14 bytes of each packet, corresponding to the Ethernet header, are entirely excluded as LDPI focuses on the IP layer and above. Within the IP header, the 4-byte source and destination IP address fields are also ignored. LDPI inspects the remainder of the IP header and continues to analyze the protocol-specific header and payload, up to a maximum of 60 bytes per packet. This includes protocol-dependent fields (e.g., TCP, UDP, or ICMP headers) and data, with the size limit ensuring efficient processing. The size of 60 bytes is a standard parameter that can be modified.

LDPI's methodology uses one-class contrastive pretraining and semi-supervised fine-tuning to boost detection capabilities. During pretraining, the model only learns on normal traffic samples. In contrast, the fine-tuning phase includes mainly normal traffic with abnormal samples. 

The pretraining phase uses the InfoNCE loss function with distribution shifts by transforming normal traffic samples to create negative examples. This method aligns with strategies proposed in \cite{tack2020csi} \cite{sohn2020learning}. A contrastive learning model built on the ResCNN architecture is employed to encode input data representations. Pretraining starts with the creation of two augmented "views" of the same input data, which are processed as paired inputs. These views undergo transformations to generate diversity, simulating variations seen in network traffic. Stochastic Gradient Descent (SGD) is used with an initial learning rate of 0.1, which is dynamically adjusted using cosine annealing after a warmup phase of 100 epochs. The warmup phase ensures a gradual increase in the learning rate to stabilize the training process.

\begin{figure*}
\centering\includegraphics[width=1\textwidth]{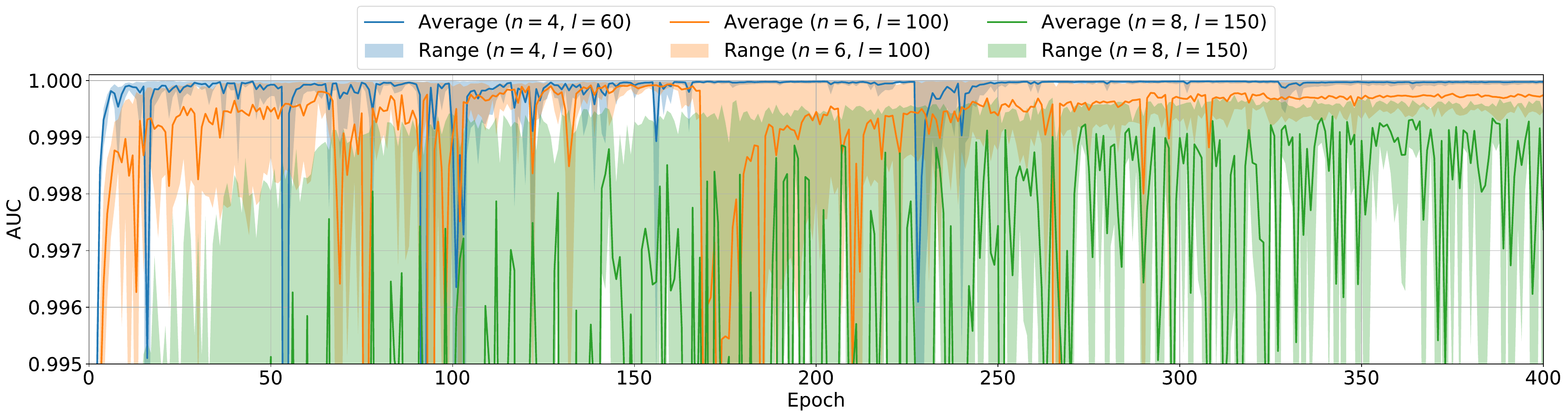}
\caption{\label{fig:comparison-auc}Comparison of AUC scores for LDPI across three training scenarios with varying parameters \(n\) and \(l\).}
\end{figure*}

\begin{table*}[ht]
\centering
\caption{Performance measurements considering the mean of the best epoch of each Training Setup}
\label{tab:performance_mean}
\begin{tabular}{@{}cccccc@{}}
\toprule
Training Setup Parameters & AUC & \multicolumn{4}{c}{F-1 Score} \\ \cmidrule(l){3-6} 
 & & \textit{ninety\_nine} & \textit{near\_max} & \textit{max} & \textit{hundred\_one} \\ \midrule
 \(n = 4\) and \(l = 60\) & 0.999994 $\pm$ 0.000003 & 0.999581 $\pm$ 0.000035 & 0.999378 $\pm$ 0.000446 & 0.999351 $\pm$ 0.000520 & 0.999351 $\pm$ 0.000520 \\ \midrule
 \(n = 6\) and \(l = 100\) & 0.999951 $\pm$ 0.000037 & 0.999287 $\pm$ 0.000229 & 0.997420 $\pm$ 0.001910 & 0.997402 $\pm$ 0.001942 & 0.997359 $\pm$ 0.001969 \\ \midrule
 \(n = 8\) and \(l = 150\) & 0.999283 $\pm$ 0.000579 & 0.990738 $\pm$ 0.007490 & 0.977132 $\pm$ 0.009367 & 0.976840 $\pm$ 0.009243 & 0.976609 $\pm$ 0.009426 \\ \midrule
\end{tabular}
\end{table*}

The model is fine-tuned on a mix of normal and a limited set of abnormal samples using the Deep SAD loss, as described in \cite{ruff2019deep}. This phase begins with preprocessing the network traffic data to ensure that features critical for anomaly detection are effectively captured. Drawing on techniques such as those in \cite{LUN23}, the data preprocessing pipeline extracts the initial packets of network flows, truncates or pads them to a fixed size, normalizes the byte values, and anonymizes sensitive fields such as MAC and IP addresses. These steps ensure consistent and privacy-preserving data input.

During training, the fine-tuning phase focuses on leveraging the learned representations from the pretraining stage to enhance anomaly detection capabilities. The Deep SAD loss minimizes the discrepancy between latent representations of normal samples, while maintaining separation from abnormal samples injected during training. 

The training process iteratively updates the model parameters using the Adam optimizer, employing a two-phase learning rate schedule: an initial "search" phase with a higher learning rate followed by a "fine-tuning" phase with a reduced rate. The model also integrates validation checkpoints to assess performance on unseen data and adjusts hyperparameters dynamically to balance precision and recall. 

The training dataset for LDPI, known as TII-SSRC-23 \cite{HER23}, was designed to address the limitations of previous datasets and better represent the complex nature of modern network threats. Considering our scenario of LDPI being deployed on Ghaf's Connection VM, we created our specific benign dataset based on normal operations within a Ghaf Application VM running the Chromium web browser.


To further refine the LDPI’s training process for optimal performance within Ghaf, we explored the impact of adjusting the parameters \(n\) (number of packets per sample) and \(l\) (size of each packet in bytes). This experiment involved three distinct scenarios to assess how changes in packet quantity and size influence the model’s ability to discern normal from malicious traffic. In the first scenario, we trained the model using 6,480 normal samples with \(n = 4\) and \(l = 60\). The second scenario adjusted the parameters to \(n = 6\) and \(l = 100\), utilizing 6,396 normal samples. Lastly, the third scenario increased the values further to \(n = 8\) and \(l = 150\), with a dataset comprising 6,202 normal samples. Figure \ref{fig:comparison-auc} shows the AUC of the LDPI model across the different training scenarios. The graph displays the average AUC (Area Under the Curve) performance obtained over 5-folds for each group. The shaded areas represent the range of AUC values observed across multiple training iterations. AUC measures the ability of a model to discriminate between classes and is used as an aggregate measure of performance across all possible classification thresholds \cite{JIN05}. An increasing AUC over time during model training suggests that the model is improving in distinguishing between the classes across all thresholds. This is generally seen as the model learns better features from the training data. Each training configuration was conducted over a consistent regimen of 2000 pretrain epochs followed by 400 training epochs. 

Table~\ref{tab:performance_mean} presents the performance metrics for the LDPI model across different training setups, each characterized by varying packet quantities (\(n\)) and sizes (\(l\)). Each line reports the mean and standard deviation (\(\pm\)) of the best epoch performance obtained over 5-folds training iterations for each configuration. These metrics include the AUC and the F-1 Score, which are critical in assessing the model's accuracy and precision. We establish various detection thresholds during the fine-tuning phase to accommodate different operational requirements and sensitivity levels. The F-1 Score, a harmonic mean of precision and recall, is presented for four pre-defined thresholds, each tailored to assess the model's performance across a spectrum of anomaly sensitivities derived from the normal traffic data in the training set:
\begin{itemize}
    \item \textit{ninety\_nine}: Set at the 99th percentile of anomaly scores, this threshold captures the upper limit of normal behavior, flagging the top 1\% of anomalous activity based on the normal distribution.
    \item \textit{near\_max}: Positioned at the 99.99th percentile, this threshold is designed to identify extreme anomalies by capturing nearly the highest scores, providing a stringent test for outlier detection.
    \item \textit{max}: Corresponds to the absolute maximum anomaly score observed in the normal dataset, serving as a critical boundary for detecting the most pronounced deviations.
    \item \textit{hundred\_one}: Calculated as 1.01 times the maximum anomaly score, this threshold tests the model's response to potential outlier values that surpass historical extremes.
\end{itemize}


\begin{table*}[t]
\centering
\caption{CPU Usage Statistics considering LDPI on Ghaf's Connection VM during \textit{Hey} tool execution}
\label{table:cpu_and_mem_usage_ldpi}
\begin{tabular}{lcccccc}
\hline
\multirow{2}{*}{\textbf{Statistic}} & \multirow{2}{*}{\textbf{No LDPI}} & \multirow{2}{*}{\textbf{With LDPI}} & \multicolumn{2}{c}{\textbf{Suricata (IDS mode)}} & \multicolumn{2}{c}{\textbf{Snort (IDS mode)}} \\
\cline{4-7}
& & & \textbf{No rules} & \textbf{ET Open rules} & \textbf{Default} & \textit{\textbf{alert\_fast}}  \\
\hline
Average CPU (\%) & 3.24 & 4.19 & 3.99 & 4.05 & 3.30 & 3.25\\
Maximum value CPU (\%) & 10.20 & 11.90 & 10.00 & 10.20 & 9.90 & 10.10  \\
Standard Deviation CPU (\%) & 2.34 & 2.85 & 2.42 & 2.33 & 1.92 & 1.95 \\
Average Memory (MB) & 355.72 & 568.92 & 409.23 & 854.61 & 461.59 & 406.95  \\
Maximum Memory (MB) & 359.20 & 575.60 & 419.70 & 862.40 & 465.40 & 420.20  \\
Standard Deviation Memory (MB) & 2.46 & 4.68 & 6.18 & 5.61 & 3.01 & 7.40  \\
\hline
\end{tabular}
\end{table*}

\section{Evaluation}
\label{sec:eva}
Our testing environment comprises a deployed instance of the Ghaf Framework on a Lenovo ThinkPad X1 Gen10 laptop equipped with a 12th Generation Intel Core i7-1270P Processor with 32 GB of memory. This configuration hosts several virtual machines, including an Admin VM, a Connection VM, a Display VM, and a set of Application VMs, specifically structured to mimic a realistic operational setting for the Ghaf Framework. LDPI was deployed at the Connection VM, which is responsible for enabling the network to be used by all the other Ghaf VMs. The Connection VM has 1024 MB of memory and one Core. Our deployed instance was connected via \textit{Wi-Fi} to the internet.


\subsection{Performance Measurements - hey tool}
\label{subsec:perf_hey}


We employ the \textit{hey}\footnote{\url{https://github.com/rakyll/hey}} tool to generate network traffic. \textit{Hey} is a minimalistic tool designed to generate load on a web application, similar in functionality to ApacheBench (ab). \textit{Hey} is employed to simulate a controlled high-traffic environment by sending requests to a predefined list of websites. This method allows us to evaluate LDPI's performance under a synthetic stress test designed to mimic a surge in network activity. The chosen websites for this test include domains such as \textit{Google}, \textit{YouTube}, \textit{Facebook}, \textit{Instagram}, \textit{Wikipedia}, \textit{Yahoo}, and \textit{tii.ae}. The performance evaluation focuses on key metrics such as CPU usage percentage and memory usage in MB. The \textit{hey} tool was configured to run over a 30-minute period, cycling through the list of chosen websites, sending 100 requests (\textit{-n 100}) to each with a concurrency level of 5 (\textit{-c 5}), simulating 5 users. Each website was engaged for a 10-second interval (\textit{-z 10s}), aiming to evenly distribute the load across the test duration and websites. 

To comprehensively evaluate LDPI's performance, we conducted this test using six different configurations on the Connection VM:

\begin{itemize}
    \item \textbf{Baseline (No LDPI):} Connection VM without LDPI.
    
    \item \textbf{LDPI Enabled:} Connection VM with LDPI enabled.
    
    \item \textbf{Suricata (No Rules):} This configuration included Suricata without any active rules to measure its base performance overhead.
    
    \item \textbf{Suricata with ET Open Ruleset:} Suricata was enabled with the Emerging Threats (ET) Open Ruleset\footnote{\url{https://rules.emergingthreats.net/OPEN_download_instructions.html}}. The ET Open Ruleset applied in our test included 54 rule files, totaling 42,059 rules, with 1,172 IP-only rules, 4,305 inspecting packet payloads, and 36,551 analyzing application-layer traffic.

    \item \textbf{Snort:} In this configuration, we enabled Snort passively monitoring network traffic without actively blocking any detected anomalies. The applied ruleset was the \textit{Snort v3.0 community ruleset}\footnote{\url{https://www.snort.org/downloads/community/snort3-community-rules.tar.gz}}, which includes a set of 625 built-in rules.

    \item \textbf{Snort (\textit{\textbf{alert\_fast}}):} In this configuration, we maintained the same Snort v3.0 community ruleset but enabled the \textit{alert\_fast} mode to optimize logging efficiency. 
\end{itemize}

The summarized results of this evaluation are presented in Table \ref{table:cpu_and_mem_usage_ldpi}. The results were obtained by leveraging the \textit{top} command in batch mode, which facilitated the capture of system metrics, including CPU and memory usage, at 5-second intervals over a span of 30 minutes. 

The results indicate that enabling LDPI, Suricata, and Snort had varying impacts on system performance. In terms of CPU usage, all configurations exhibited a small to moderate increase compared to the baseline (no LDPI). LDPI resulted in an average CPU usage of 4.19\%, representing a 29.3\% increase over the baseline (3.24\%). Snort demonstrated a lower CPU overhead, a 1.85\% increase over the baseline, and 0.31\% when configured with \textit{alert\_fast} mode. Memory consumption exhibited notable variations across the tested configurations. LDPI had an average memory usage of 568.92 MB, reflecting a 59.9\% increase compared to the baseline (355.72 MB). Suricata with the ET Open ruleset had the highest memory footprint, averaging 854.61 MB, which is 140.3\% higher than the baseline. Snort’s memory consumption fell between LDPI and Suricata.

\subsection{Performance Measurements - Flooding Attacks}
We evaluated LDPI’s performance under active flood attacks targeting Ghaf’s Connection VM. The test environment included a network router with multiple devices connected, among which was the Ghaf laptop running LDPI on its Connection VM. We simulated diverse kinds of flooding attacks (i.e., DNS, HTTP, UDP, and SYN). All tests were conducted over a 15-minute period, allowing us to observe LDPI’s response and resource consumption under distinct flooding scenarios. Each test was performed twice: (i) without LDPI enabled on the Connection VM and (ii) with LDPI enabled. We additionally ran a \textit{Base} test to gather idle Connection VM performance without a flooding attack happening. Data was collected using the \textit{top} command in batch mode, capturing CPU and memory usage at 5-second intervals. 

\begin{table}[t]
\centering
\caption{Comparison of CPU and Memory Usage for LDPI on Ghaf's Connection VM during flooding attacks}
\label{table:cpu_mem_usage_ldpi_comparison}
\begin{tabular}{lccccc}
\hline
\multicolumn{6}{c}{\textbf{Connection VM - no LDPI}}\\
\hline
\textbf{Flooding Attack} & \textbf{DNS} & \textbf{HTTP} & \textbf{UDP} & \textbf{SYN} & \textbf{Base} \\
\hline
Avg CPU (\%) & 12.86 & 91.26 & 10.17 & 48.85 & 0.76 \\
Max CPU (\%) & 21.10 & 98.70 & 12.70 & 64.40 & 16.60 \\
Std Dev CPU (\%) & 3.02 & 17.99 & 1.02 & 9.32 & 1.20 \\ 
Avg Mem (MB) & 329.81 & 872.03 & 323.01 & 435.84 & 323.52 \\
Max Mem (MB) & 330.80 & 966.70 & 323.20 & 436.40 & 323.90 \\
Std Dev Mem (MB) & 0.56 & 132.89 & 0.10 & 0.70 & 0.13 \\
\hline
\multicolumn{6}{c}{\textbf{Connection VM - with LDPI}}\\
\hline
\textbf{Flooding Attack} & \textbf{DNS} & \textbf{HTTP} & \textbf{UDP} & \textbf{SYN}  & \textbf{Base} \\
\hline
Avg CPU (\%)  & 0.89 & 1.28 & 9.15 & 9.82 & 0.90 \\
Max CPU (\%) & 9.10 & 15.40 & 11.20 & 28.00 & 9.10 \\
Std Dev CPU (\%) & 0.66 & 1.26 & 0.89 & 2.16 & 0.72 \\ 
Avg Mem (MB) & 517.69 & 565.26 & 525.29 & 625.59 & 517.29 \\
Max Mem (MB) & 519.60 & 571.30 & 526.20 & 635.50 & 522.40 \\
Std Dev Mem (MB) & 0.69 & 4.23 & 1.61 & 9.91 & 2.69 \\
\hline
\end{tabular}
\end{table}

Table \ref{table:cpu_mem_usage_ldpi_comparison} presents a summary of CPU and memory usage statistics during the flooding attacks. It is important to note that in all tested flooding attack scenarios, LDPI successfully detected the attacks and blocked the source IP at the onset of each attack.

In terms of memory consumption, the worst-case (HTTP) flood, LDPI reduces average memory consumption by approximately 35\% relative to running without LDPI, indicating effective suppression of attack-driven allocation. This comes with an idle cost of about 194 MB on the Connection VM, which can outweigh the avoided growth despite early blocking. 

LDPI substantially reduces CPU consumption during active flooding by detecting and blocking sources at onset. Relative to running without LDPI, average CPU usage drops by nearly 99\% under HTTP floods, by roughly 93\% under DNS, and by about 80\% under SYN. UDP shows a modest reduction of around 10\%.

\section{Conclusion}
\label{sec:conc}

This paper presented the Lightweight Deep Anomaly Detection for Network Traffic (LDPI) module within the Ghaf framework for edge intrusion detection, and compared it with Suricata and Snort on the same platform. Suricata with the ET Open ruleset raised memory usage by up to 140\% over baseline, whereas LDPI kept a more controlled footprint. Snort used less memory than LDPI, reflecting that rule-based IDSes can be lighter depending on rule count, though with reduced coverage for unknown threats. Under flooding, LDPI cut the average CPU by about 70\% across attacks (nearly 99\% in HTTP). Finally, our learning pipeline-one-class contrastive pretraining on benign traffic followed by semi-supervised fine-tuning (Deep SAD) with a small anomaly set-achieved high discriminative power (AUC \(\approx 0.999\) across 5 folds) with modest runtime overhead on a laptop-class edge node.

Suricata and Snort are signature-driven engines whose effectiveness depends on rule coverage and freshness, and their memory footprint typically grows as rulesets expand. LDPI, by contrast, is anomaly-oriented, providing a complementary defense for emerging or previously unseen threats. A practical deployment could utilize LDPI as a first-stage filter, inspecting only the early bytes of each flow, escalating only the most anomalous slice of traffic (e.g., the top \(\sim 10\%\)) to a rule-based engine for full-packet deep inspection. This cascade would preserve the high recall of anomaly detection while keeping CPU and memory costs controlled.

\bibliographystyle{IEEEtran}
\bibliography{IEEEabrv,bib}

\end{document}